\documentclass[twocolumn,prl,showpacs]{revtex4}
\usepackage{amsmath,amsthm}
\usepackage{epsfig}
\usepackage{graphics}
\usepackage{graphicx}
\usepackage{dcolumn}
\usepackage{bm}

\begin{document}

\title{Charge dynamics in the normal state of the iron oxypnictide superconductor LaFePO}
\author{M. M. Qazilbash$^{\ast}$, J. J. Hamlin, R. E. Baumbach, M. B. Maple, and D. N. Basov}
\affiliation{Physics Department, University of California-San
Diego, La Jolla, California 92093}

\date{\today}

\begin{abstract}
We present the first infrared and optical study in the normal
state of \textit{ab}-plane oriented single crystals of the
iron-oxypnictide superconductor LaFePO. We find that this material
is a low carrier density metal with a moderate level of
correlations and exhibits signatures of electron-boson coupling.
The data is consistent with the presence of coherent
quasiparticles in LaFePO.
\end{abstract}


\maketitle

The recent discovery of superconductivity in the iron
oxy-pnictides promises to be an important milestone in the field
of superconductivity \cite{hosono1,hosono2}. Here is a new class
of quasi-two-dimensional materials with a layered structure and
relatively high superconducting \textit{T}$_c$ values
\cite{hosono2,ren} rivalling the doped superconducting cuprates.
Electronic conduction is believed to occur in the iron-pnictogen
layers \cite{hosono3}, similar to the cuprates where the charge
carriers are delocalized in the copper-oxygen planes. While
maintaining the cuprates as the benchmark for judging the new
family of iron-based superconductors, it is worthwhile to note
that the high superconducting transition temperatures are not the
only puzzling features of the cuprates. Their normal state
properties are unconventional and include pseudogap features,
strong-coupling effects, charge-spin self-organization, and bad
metal behavior \cite{basovreview}. Superconductivity arises from
the Cooper instability of this anomalous metallic state of the
cuprates. It is therefore imperative to investigate and understand
the normal state of the iron-oxypnictide superconductors,
especially their in-plane properties. Access to in-plane
properties without contamination by the inter-plane response is
only possible by studying single crystals.

In this Letter, we report on the infrared and optical response in
the normal state of single crystals of LaFePO, the first
superconducting iron-oxypnictide to be discovered \cite{hosono1}.
The \textit{ab}-plane infrared data allow us to examine the
dynamical properties of the charge carriers in the iron-pnictogen
layers. We find that LaFePO is a low-carrier density, moderately
correlated metal with well-defined quasiparticles. Moreover, we
find evidence of electron-boson coupling in the optical
quasiparticle self-energy, demonstrating that many-body effects
cannot be neglected.

The LaFePO crystals were grown by a flux method whose details are
given in Ref.\onlinecite{maple}. Resistivity and magnetization
measurements reveal a superconducting \textit{T}$_c$ of $\approx$
6 K, with complete Meissner shielding in the superconducting
state. The crystals are platelets, typically 0.5 mm $\times$ 0.5
mm $\times$ 0.05 mm in size. The \textit{ab}-plane reflectance was
measured in the near-normal incidence geometry in a Bruker v66
Fourier Transform Infrared Spectrometer at frequencies between 100
cm$^{-1}$ and 24000 cm$^{-1}$. The reflectance measurements were
performed at the following temperatures: 298 K, 250 K, 200 K, 150
K, 100 K, 50 K, and 10 K. We obtained the optical constants
through fitting the reflectance data with Drude and Lorentzian
oscillators, and Kramers-Kronig constrained variational dielectric
functions as described in Ref.\onlinecite{kkvariational}. In
addition, variable-angle spectroscopic ellipsometry was performed
in the frequency range 5500 cm$^{-1}$ and 25000 cm$^{-1}$ which
improves the accuracy of the extracted optical constants in this
frequency range.

\begin{figure}[t]
\epsfig{figure=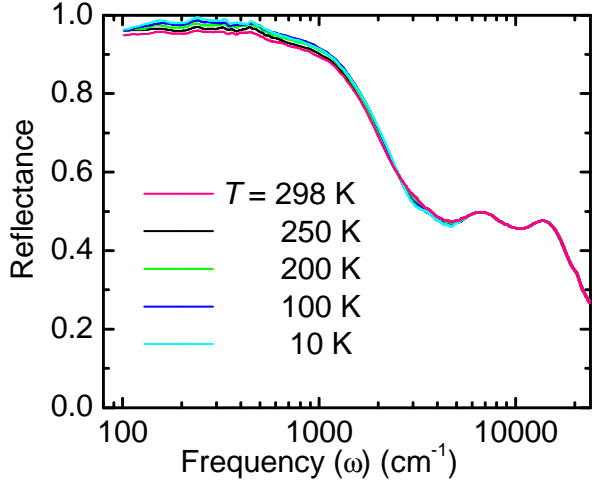,width=80mm,height=65mm}
\caption{(color online): Plots of \textit{ab}-plane reflectance of
single crystal LaFePO as a function of frequency for several
representative temperatures.} \label{reflectance}
\end{figure}

\begin{figure}[t]
\epsfig{figure=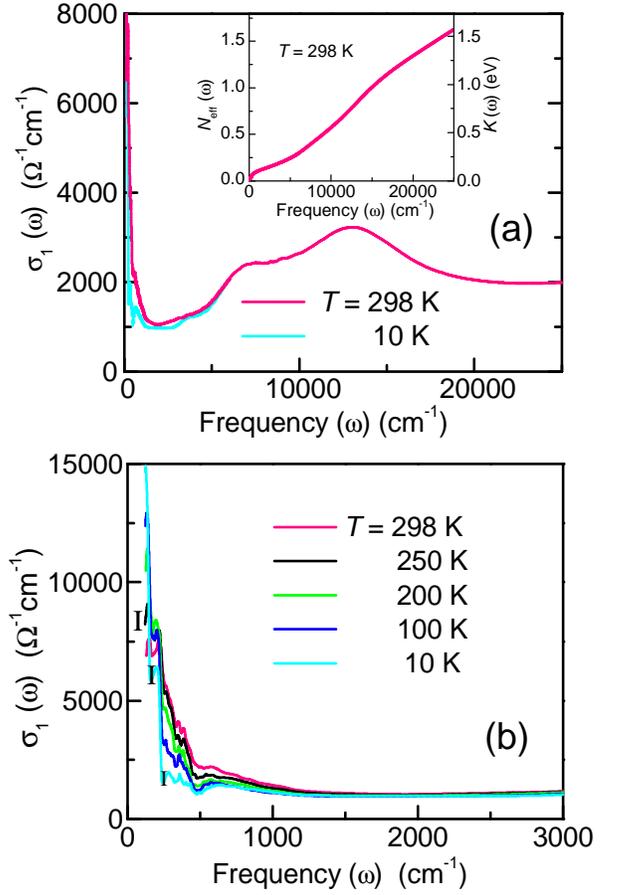,width=80mm,height=120mm}
\caption{(color online): (a) Real part of the \textit{ab}-plane
optical conductivity $\sigma_1$($\omega$) of LaFePO is plotted as
a function of frequency for \textit{T} = 298 K and \textit{T} = 10
K. The inset in panel (a) shows the frequency dependence of the
effective number of charge carriers per formula unit
\textit{N}$_{eff}$($\omega$) and the energy \textit{K}($\omega$)
for \textit{T} = 298 K. (b) Systematic temperature dependence of
$\sigma_1$($\omega$). The uncertainty in $\sigma_1$($\omega$) in
panel (b) is less than or equal to the thickness of the lines for
$\omega$ $>$ 350 cm$^{-1}$. However, the uncertainty in
$\sigma_1$($\omega$) increases at lower frequencies as indicated
by the error bars.}\label{sigma1}
\end{figure}

The temperature dependence of the \textit{ab}-plane reflectance of
LaFePO crystals is displayed in Fig.\ref{reflectance}. This
material is a very good metal as attested by the high values of
the reflectance ($\geq$ 95\%) at low frequencies. Two prominent
hump-like features between 5000 cm$^{-1}$ and 15000 cm$^{-1}$ are
likely due to interband transitions. In Fig.\ref{sigma1}, we
present the real part of the optical conductivity
$\sigma_1$($\omega$) over a broad frequency range. There is a
clear Drude-like feature at low frequencies followed by two
prominent inter-band transitions centered at 7100 cm$^{-1}$ and
13000 cm$^{-1}$. A weaker interband transition centered at 3700
cm$^{-1}$ becomes more evident in the $\sigma_1$($\omega$) data
obtained at \textit{T} = 10 K. On the basis of band-structure
calculations, we assign these interband transitions to the bands
formed by the 3\textit{d} orbitals of Fe
\cite{lebegue,che,georges}. Interestingly, the Drude part of the
conductivity that represents intra-band excitations is
well-separated from the higher-lying inter-band transitions. Such
a separation is not usually evident in strongly correlated metals,
specifically in most cuprates \cite{basovreview}. The difficulty
in separating the Drude part from the interband transitions in the
cuprates arises from the strong incoherent contribution to the
conductivity in the mid-infrared frequency range
\cite{basovreview}.

A plasma frequency $\omega_p$ of 14900 cm$^{-1}$  for the
delocalized charge carriers is obtained by integrating the low
frequency part of $\sigma_1$($\omega$): $\omega_p^2 =
8\int_{0}^{\omega_c}\sigma_1(\omega)d\omega$. A cut-off frequency
$\omega_c$ = 3000 cm$^{-1}$ was chosen above which interband
transitions begin to dominate the optical response of this system.
The frequency-dependence of the spectral weight associated with
$\sigma_1$($\omega$) is plotted in the inset of Fig.\ref{sigma1}
in two ways: as the effective number of charge carriers per
formula unit \textit{N}$_{eff}$($\omega$) participating in the
optical transitions \cite{basovreview}, and as an energy
\textit{K}($\omega$) \cite{millisndoped}.

\begin{align}
N_{eff}(\omega)=\frac{2m_eV}{\pi{e^2}}\int_{0}^{\omega}{\sigma_1(\omega'}){d\omega'}\\
K(\omega)=\frac{{\hbar}c_0}{e^2}\int_{0}^{\omega}\frac{2\hbar}{\pi}{\sigma_1(\omega'}){d\omega'}
\label{energy}
\end{align}

In the above equations, \textit{m}$_e$ is the bare electron mass,
\textit{V} is the volume of the unit cell per formula unit, and
\textit{c}$_0$ is the \textit{c}-axis lattice constant. Assuming
an intraband cutoff frequency of 3000 cm$^{-1}$, we estimate 0.16
itinerant charge carriers per formula unit. The energy
\textit{K}($\omega$) at the intraband cutoff $\omega_c$ = 3000
cm$^{-1}$ is simply the kinetic energy of the itinerant carriers
and is equal to 0.15 eV. Correlations effects are expected to
reduce the kinetic energy compared to the band theory value
\cite{millisndoped,millisnatphys}. Therefore, it would be useful
to compare the experimental value with future band structure
calculations for a quantitative estimate of the strength of
correlations. In this context, one ought to mention that the
kinetic energy in the cuprates near optimal doping is about 0.1 eV
\cite{millisnatphys} and this is generally considered as evidence
of strong correlations. The kinetic energy in LaFePO is higher
than in optimally-doped cuprates, even though the effective number
of itinerant carriers is similar \cite{uchida}, indicating lesser
degree of correlations in LaFePO.

\begin{figure}[t]
\epsfig{figure=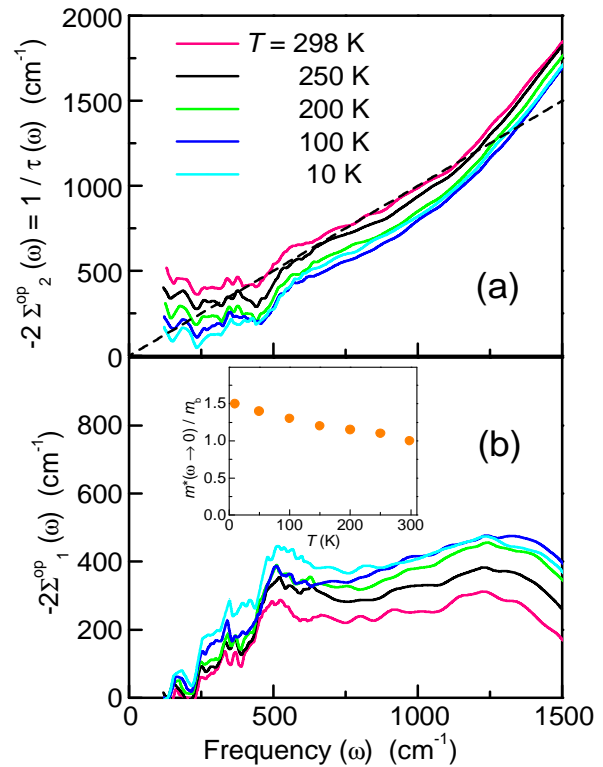,width=80mm,height=105mm}
\caption{(color online) Panels (a) and (b) respectively display
the frequency dependence of the imaginary part of the optical
quasiparticle self energy -2$\Sigma^{op}_2$($\omega$) [or
scattering rate 1/$\tau$($\omega$)], and the real part of the
optical quasiparticle self- energy -2$\Sigma^{op}_1$($\omega$) for
representative temperatures in the normal state of LaFePO.  The
dashed line in panel (a) represents the equation $\omega\tau$ = 1.
The inset in panel (b) shows the temperature dependence of the
mass enhancement factor in the low frequency limit
\textit{m}*($\omega$ $\rightarrow$ 0)/\textit{m}$_b$.}
\label{xtendedDrude}
\end{figure}

We plot the temperature dependence of the low-frequency part of
$\sigma_1$($\omega$) in Fig.\ref{sigma1}b. The Drude-like peak at
low frequencies becomes sharper and the conductivity at the lowest
measured frequencies increases with decreasing temperature. The
spectral weight is transferred from high frequencies to low
frequencies with decreasing temperature, presumably below our
low-frequency experimental cutoff. We also clearly observe an
anomaly in $\sigma_1$($\omega$) near 500 cm$^{-1}$ which becomes
more prominent as the temperature is lowered. In order to gain
more insight into this feature and to obtain the dynamical
behavior of the charge carriers, we performed the extended Drude
analysis \cite{basovreview,timusk} on the real and imaginary parts
of the optical conductivity [$\sigma_1$($\omega$),
$\sigma_2$($\omega$)]. Via the extended Drude analysis, we extract
the real and imaginary parts of the optical quasiparticle
self-energy [$\Sigma^{op}_1$($\omega$),$\Sigma^{op}_2$($\omega$)]
which we present in Fig.\ref{xtendedDrude}a,b. The effects of
many-body interactions are encoded in
$\Sigma^{op}_1$($\omega$),$\Sigma^{op}_2$($\omega$). The
self-energy is related to the optical conductivity, the scattering
rate 1/$\tau$($\omega$) and the mass enhancement factor
\textit{m}*($\omega$)/\textit{m}$_b$ (\textit{m}$_b$ is the band
mass) by the following equations:

\begin{align}
-2\Sigma^{op}_2(\omega)=\frac{1}{\tau(\omega)}
=\frac{\omega_p^2}{4\pi}\Big(\frac{\sigma_{1}(\omega)}{\sigma_{1}^2(\omega)+\sigma_{2}^2(\omega)}\Big)\\
1-\frac{2}{\omega}\Sigma^{op}_1(\omega)=\frac{m^*(\omega)}{m_b}=
\frac{\omega_p^2}{4\pi\omega}\Big(\frac{\sigma_{2}(\omega)}{\sigma_{1}^2(\omega)+\sigma_{2}^2(\omega)}\Big)
\label{selfenergy}
\end{align}

We note that in a system like LaFePO in which multiple bands cross
the Fermi energy, the self-energy deduced from the infrared data
is to be regarded as an average of the contributions from the
relevant bands.

The anomaly in $\sigma_1$($\omega$) appears as a kink in the
scattering rate at $\approx$ 500 cm$^{-1}$ for all measured
temperatures between 298 K and 10 K. This kink has a weak
temperature dependence - it becomes only marginally more
pronounced at lower temperatures and the frequency of the onset of
the upturn in 1/$\tau$($\omega$) increases from 450 cm$^{-1}$ at
\textit{T} = 298 K to 460 cm$^{-1}$ at \textit{T} = 10 K. The kink
in the scattering rate appears in the real part of the self-energy
as a peak centered at $\approx$ 500 cm$^{-1}$ (62 meV) seen in
Fig.\ref{xtendedDrude}b. The most probable explanation of the kink
in the scattering rate and the peak in real part of the self
energy is in terms of the coupling of charge carriers to a bosonic
mode \cite{basovreview,timusk}. Infrared data alone cannot
determine whether the bosonic mode is of phononic or magnetic
origin. The kink in 1/$\tau$($\omega$) may, alternatively, be due
to a pseudogap \cite{basovreview}. However, this latter
interpretation would be inconsistent with the recent Angle
Resolved Photoemission Spectroscopy (ARPES) data on single
crystals of LaFePO where the pseudogap is not seen \cite{zxshen}.
The electron-boson coupling features identified in our infrared
studies warrant further investigation by other spectroscopic
probes, for example, ARPES with higher energy resolution,
tunneling spectroscopy and neutron scattering. We note here that
ARPES data on single crystals of the related iron-arsenide
superconductor (Sr/Ba)$_{1-x}$K$_x$Fe$_2$As$_2$ reveals kinks in
the dispersion (energy range 15-50 meV) which appear to be related
to the energy scales of magnetic excitations \cite{mzhasan}.

Above the kink feature, the scattering rate increases
monotonically with frequency. Electron-phonon scattering theory
predicts a constant, frequency-independent scattering rate above
the cutoff of the phonon spectrum \cite{basovreview,VO2}.
Calculations of the phonon spectrum for LaFePO are not yet
available. However, the cutoff is not likely to exceed 600
cm$^{-1}$ inferred from calculations on the iron oxy-arsenides
\cite{djsingh,boeri}. Therefore, the absence of saturation of
1/$\tau$($\omega$) in LaFePO suggests that in addition to
electron-phonon scattering, there are other scattering channel(s)
presumably arising from electronic correlations. The available
experimental data on LaFePO suggests this material is close to a
spin density wave (SDW) instability \cite{dHvA} although there is
no evidence yet for an SDW ground state \cite{uemura}. We note
that partial nesting of the Fermi surface \cite{zxshen,dHvA}
and/or coupling to magnetic excitations \cite{uemura,mzhasan} may
explain the monotonic increase of the frequency-dependent
scattering rate of the quasiparticles. The scattering rate for
LaFePO at the lowest measured temperature \textit{T} = 10 K is
close to or below the line $\omega\tau$ = 1 at low frequencies
(see Fig.\ref{xtendedDrude}a). Within conventional Fermi Liquid
Theory, the $\omega\tau$ = 1 line separates well-defined
quasiparticles from incoherent excitations \cite{basovreview,VO2}.
The data imply that well-defined quasiparticle excitations exist
in LaFePO, consistent with the observation of deHaas-vanAlphen
oscillations in all the Fermi surface pockets \cite{dHvA}.

The mass enhancement factor ($\omega$ $\rightarrow$ 0) at
\textit{T} = 10 K is 1.5 $\pm$ 0.1 which is in reasonably good
agreement with the factor of nearly 2 obtained from ARPES data and
deHaas-vanAlphen measurements \cite{zxshen,dHvA}. The rather
modest mass enhancement factor is only weakly dependent on
temperature (see inset of Fig.\ref{xtendedDrude}b) and these
observations together suggest a moderate degree of electronic
correlations.

It is instructive to compare the normal state infrared response of
superconducting LaFePO to that of MgB$_2$ and the superconducting
cuprates. First, the plasma frequency of LaFePO is comparable to
the values seen in doped superconducting cuprates
\cite{basovreview}, indicating similar carrier densities in the
two systems. This is in contrast to MgB$_2$, a high carrier
density system, where the plasma frequency is nearly 48000
cm$^{-1}$ ($\approx$ 6 eV) \cite{mgb2}. A linear frequency
dependent scattering rate is seen in many of the cuprates near
optimal doping and its origin still remains a mystery
\cite{basovreview}. However, the 1/$\tau$($\omega$) data in LaFePO
does not show obvious signs of a linear frequency dependence. The
scattering rate data extracted from infrared measurements on
under-doped and optimally-doped  cuprates lie above the
$\omega\tau$ = 1 line and suggests predominance of incoherent
excitations \cite{basovreview} over a wide frequency range. This
is to be contrasted with the predominance of quasiparticle
excitations in LaFePO whose 1/$\tau$($\omega$) data resembles what
is seen in the over-doped cuprates which are considered closer to
the Fermi Liquid regime \cite{basovreview}. In MgB$_2$, the
scattering rate of charge carriers is significant, mainly due to
strong electron-phonon coupling \cite{mgb2}. Hence, we infer that
many-body interactions in LaFePO are weaker than in the
under-doped and optimally-doped cuprates and MgB$_2$, but cannot
be considered negligible.

The cuprates, especially low-doped ones, exhibit bad metal
behavior \cite{kivelson} wherein the resistivity exceeds the
Ioffe-Regel-Mott (IRM) limit of metallic transport and shows no
signs of saturation at high temperatures. The IRM limit is
generally expressed as \textit{k}$_F$\textit{l} $\approx$ 1
(\textit{k}$_F$ is the Fermi wavevector and \textit{l} is the
electronic mean free path) or \textit{l} $\approx$ \textit{a}
(\textit{a} is the in-plane lattice constant). The IRM limit is
regarded as a regime where the quasiparticle description is not
valid \cite{basovreview,VO2,kivelson}. At room temperature, the
in-plane resistivity ($\rho_{ab}$) of LaFePO yields
\textit{k}$_F$\textit{l} $\approx$ 2 and this value increases by a
factor of 25 just above \textit{T}$_c$ \cite{maple,footnote1}. At
least at low temperatures, the \textit{k}$_F$\textit{l} value is
well above the IRM limit and therefore indicates that the
quasiparticle description is applicable in LaFePO. In this
context, it would be interesting to investigate the behavior of
the resistivity of LaFePO at elevated temperatures.

The picture that emerges for LaFePO is of a low carrier density
material with well-defined quasiparticle excitations leading to
robust metallicity. Electronic correlations, though present, are
weaker than in the optimally-doped cuprates. Moreover, there is
evidence of electron-boson coupling in LaFePO. Hence, we classify
LaFePO as a moderately correlated metal. It would be interesting
to study the infrared properties of single crystals of the
iron-arsenide compounds for comparison with our LaFePO results,
especially the trends in the degree of electronic correlations.

The authors wish to thank O. Shpyrko for discussions, and A.
Kuzmenko for assistance with the software for infrared data
analysis. This work was supported in part by Department of Energy
Grant No.DE-FG03-00ER45799.


\begin{references}
\bibitem{footnote1} $\rho_{ab}$ and $k_Fl$ for a two-dimensional system are related by
$k_Fl=\frac{2\pi\hbar{c_0}}{e^2\rho_{ab}}$ for a single Fermi
surface. For LaFePO with multiple Fermi surface sheets, the $k_Fl$
value in the above equation will be reduced by a factor of about
2.
\end{references}
\end{document}